\documentclass[aps,showpacs,twocolumn,floatfix]{revtex4} 
\usepackage{graphicx}

\begin{document}
   
 
\title{Stabilization of bright solitons and vortex solitons
           in a trapless three-dimensional Bose-Einstein condensate
           by temporal modulation of the scattering length
}
\author{Sadhan K. Adhikari}
\address{Instituto de F\'{\i}sica Te\'orica, Universidade Estadual
Paulista,  01.405-900 S\~ao Paulo, S\~ao Paulo, Brazil}

\date{Received 14 November 2003; published 22 June 2004}

\begin{abstract}

Using variational and numerical solutions of the mean-field
Gross-Pitaevskii equation 
we show that a bright soliton can be stabilized in a
trapless
three-dimensional attractive Bose-Einstein condensate (BEC)  by a rapid
periodic temporal modulation of scattering length alone by using a
Feshbach resonance.  This scheme also stabilizes a rotating vortex soliton
in two dimensions.  Apart from possible experimental application in BEC,
the present study suggests that the spatiotemporal solitons of nonlinear
optics in three dimensions can also be stabilized in a layered Kerr medium
with sign-changing nonlinearity along the propagation direction.

\end{abstract}
\maketitle
 
\section{introduction}
 
Solitons are  solutions of  wave equation where
localization is obtained due to a nonlinear
attractive interaction. Solitons
 have been noted in  optics \cite{0},
high-energy physics and water waves \cite{1}, and more recently in
Bose-Einstein
condensates (BEC) \cite{2,3}. The 
bright
solitons of attractive BEC represent  local maxima \cite{3,4a0,4a1},
whereas 
dark solitons of repulsive BEC
represent local
minima \cite{2}.  

 A
classic text-book example of soliton appears  in the  following
one-dimensional
nonlinear free Schr\"odinger equation in
dimensionless units \cite{1,7}
\begin{equation} \label{1}
\left[-i
 \frac{\partial }{\partial t}
-  \frac    {\partial^2 }{\partial y^2} -
| \Psi(y,t)|^2 \right]
\Psi(y,t)  =0.
\end{equation} 
The solitons of this equation  are localized solution due to the
attractive
nonlinear
interaction $-| \Psi(y,t)|^2 $
with wave function  at time
$t$ and position
$y$:  $\Psi(y,t)= \sqrt{2|{\cal E}| }\exp(-i{\cal E} t){\mbox{sech}}
(y\sqrt
{|\cal E|})$, with $\cal E$ the energy \cite{7}. 
The Schr\"odinger equation
with a nonlinear interaction $-|\Psi|^2$ does not sustain a localized
solitonic solution in two (2D) or three dimensions (3D). However, a
radially trapped
and
axially free version
of this equation in 3D does sustain such a bright solitonic
solution \cite{4a0,4a1} which has been observed experimentally in BEC
\cite{3}.

To generate 
a soliton without  trap the repulsive
kinetic pressure has to balance  the attractive force. For a condensate of
size $L$, kinetic energy is proportional to $L^{-2}$ whereas attraction is
proportional to $L^{-D}$ in $D$ dimensions. The effective potential, which
is a sum of these two terms, has a confining minimum only for $D=1$
leading to a stable bound state \cite{ueda,8}. 
Thus no stabilization can be
obtained in
2D or 3D and any attempt to create a soliton either leads
to collapse or an expansion to infinity. 

A scheme of stabilization of a soliton in two dimensions   has been
suggested \cite{ueda,abdul}
recently by a rapid periodic temporal modulation 
of scattering
length $a$ of angular frequency $\omega$ via $a \to a_0[1-c\sin(\omega
t)]$ where $a_0$ and $c$ are
constants and $t$ is time. 
Such a modulation of the scattering length is
possible by manipulating an external magnetic field near a Feshbach
resonance \cite{4a} and has been employed in different studies of the 
BEC \cite{td}.  By such a modulation the atomic interaction
can be
easily switched between attractive and repulsive thus resulting in a rapid
contraction and expansion of the condensate. If the constants of
modulation are appropriately chosen this leads to a stabilization of the
condensate in 2D with breathing oscillation
\cite{ueda}. However, it has
been ``proved"  by
analytic and numerical calculations 
in
Ref. \cite{abdul} 
that such a stabilization does not take place in 3D.

As the problem of stabilization of a soliton in a trapless condensate  
is  of utmost interest in several areas, e.g.,
optics \cite{1,mal}, nonlinear physics \cite{1} and Bose-Einstein
condensate, we revisit this
problem and find that a temporal modification of the
scattering length
can also lead to a stabilization of the trapless soliton in 3D. We use
both variational as well as numerical solutions of the mean-field
time-dependent Gross-Pitaevskii (GP) equation to establish our claim. To
the best of our knowledge 
this is the  first suggestion of  stabilization of a trapless soliton
in 3D. We find numerically that an untrapped attractive condensate can
maintain a reasonably constant spatial profile over a large interval of
time  through the temporal modulation of the s-wave
scattering length. We also point out a possible reason for the failure to
stabilize a bright soliton in 3D in Ref. \cite{abdul}.

The present approach is also  extended  
to   stabilize   vortex solitons in 2D \cite{9} with
angular momentum
$\hbar$
per atom  in  the axial (azimuthal) direction. 
The two dimensional geometry can be achieved in an
axially symmetric configuration 
by applying
a strong trap in the axial direction \cite{ueda}. This is also equivalent
to applying a very weak trap in the radial direction. In both cases the
radial dimension of the condensate is much larger than the axial dimension
and a 2D treatment can be justified.  Vortex solitons are rotating
solitons of an attractive condensate and it is suggestive  that a similar 
scheme can also be used to stabilize a vortex soliton of a
trapless condensate 
in 
3D. After the experimental observation
\cite{6a}  of a
vortex   in a rotating BEC and theoretical prediction \cite{4a1} of a
radially trapped 
bright
vortex
soliton, 
the experimental stabilization  of
trapless vortex
solitons  seems viable.  
However, the stabilization of a vortex
soliton in 3D  calls for  a full 
three-dimensional
calculation and is beyond the scope of the present study.
In the present study we stick to a two-dimensional circularly 
symmetric configuration. 
It is well-known that in a real 3D 
problem, the vortices are unstable against azimuthal perturbation which
breaks the azimuthal symmetry and calls for a full 3D
treatment. However, these degrees of   freedom are expected to be
partially suppressed in the limit
of a very strong azimuthal trap or a very weak radial trap 
when the vortex dynamics  becomes
essentially two dimensional. In that limit a circularly symmetric 2D
calculation for a bright vortex soliton should be sufficient. 
In this paper we find that such a 2D vortex
is stable against radial perturbation. However, we have not established
its stability under transverse perturbation. The stability under
transverse perturbation can be tested by a calculation in cartesian
coordinates and
would be an investigation of future interest.

In Sec. II we present the mean-field model which we use in our study. 
In Sec. III and IV, respectively,  we present the variational and
numerical 
results of our investigation  and in Sec. V we
present our conclusions.

\section{Mean-field model}

We use the mean-field GP equation for the present
investigation \cite{8}.
In terms of an external  reference angular frequency $\Omega$ and length 
$l\equiv \sqrt {\hbar/(m\Omega)}$ the GP equation for the 
time-dependent Bose-Einstein condensate wave
function $\Psi({\vec r};t)$ at position ${\vec r}$ and time $t $ can be
rewritten in dimensionless form  as \cite{9}
\begin{eqnarray}\label{d1} 
\biggr[&-& i\frac{\partial
}{\partial t} -{\nabla_r^2   }
+\frac{1}{4}\left(\frac{\omega_\rho^2}{\Omega^2}\rho^2
+\frac{\omega_z^2}{\Omega^2}
z^2\right)  \nonumber \\ &+&
  8\sqrt 2 \pi n\left|
{\Psi({\vec r};t)}\right|^2
 \biggr]\Psi({\vec  r};t)=0, 
\end{eqnarray}
where 
length, time, $\nabla^2$, and wave function are expressed in units of
$l/\sqrt 2$, $\Omega^{-1}$, $(l/\sqrt 2)^{-2}$,
and $(l/\sqrt 2)^{-3/2}$, respectively. 
Here nonlinearity $n=Na/l$,   $m$
is
the mass and  $N$ the number of atoms in the
condensate, and
$a$ the atomic scattering length. The scattering length $a$ and
nonlinearity $n$ are negative for an attractive condensate and positive
for a repulsive condensate.  In Eq. (\ref{d1}) there is an axially
symmetric 
harmonic trap with 
angular frequency  $\omega_\rho$ 
in the radial direction $\rho$ and
$ \omega_z$  in  the
axial direction $z$. 
The normalization condition in  Eq. (\ref{d1}) is
$ \int d{\vec r} |\Psi({\vec  r};t)|^2 = 1. $

The quasi 2D limit of Eq. (\ref{d1}) is achieved by considering
$\Omega=\omega_z  >>\omega_\rho $. This condition is satisfied by 
taking  the limit $
\omega_\rho \to 0$ for a fixed $\Omega=\omega_z$. This corresponds to a
pancake-shaped condensate and we look for solution of the form $\Psi({\vec
r};t)= {\cal A}(z) \psi({\vec \rho};t)$ with ${\cal A}(z)$ satisfying the
one-dimensional ground-state oscillator equation
\begin{equation}
 -\frac{d^2}{dz^2} {\cal A}(z)+\frac{z^2}{4} {\cal A}(z)  
= \frac{1}{2}
{\cal A}(z),
\end{equation}
with
$|{\cal A}(z)|^2= \sqrt{{1}/(2\pi)} \exp[-z^2
/2]$ and
$\int_{-\infty}^\infty |{\cal A}(z)|^2 dz = 1$. Multiplying
Eq. (\ref{d1}) by
${\cal A}^*(z)$ and integrating over $z$ we get the quasi two-dimensional
GP
equation for $\psi({\vec \rho};t)$: \cite{ueda}
\begin{equation}\label{d2}
\biggr[-i\frac{\partial
}{\partial t} -{\nabla_\rho^2   }
+\frac{1}{4}\frac{\omega_\rho^2}{\Omega^2}\rho^2  +
  4n \sqrt{2\pi}  \left|
{\psi({\vec  \rho};t)}\right|^2
 \biggr]\psi({\vec  \rho};t)=0,
\end{equation}
with normalization $\int d\vec \rho |\psi({\vec \rho};t)|^2 =1$.
In a quantized vortex state \cite{9}, with each atom having angular
momentum $L\hbar$ along the $z$ axis, $\psi({\vec \rho}, t)=
\varphi(\rho,t)\exp (iL\phi) $ where $\phi$ is the azimuthal
angle. Then   the radial part of the GP equation 
(\ref{d2})
becomes \cite{9} \begin{eqnarray}\label{d3} &\biggr[&-i\frac{\partial
}{\partial t} -\frac{\partial^2}{\partial
\rho^2}-\frac{1}{\rho}\frac{\partial}{\partial \rho}
 +\frac{1}{4} \frac{\omega_\rho^2}{\Omega^2}\rho^2d(t)+ {L^2\over \rho^2} 
\nonumber \\ &+&
 4n\sqrt {2 \pi} \left|
{\varphi({\rho};t)}\right|^2
 \biggr]\varphi({\rho};t)=0,
\end{eqnarray}
with normalization $2\pi \int_0^\infty d\rho \rho
|\varphi({\rho};t)|^2 =1.$  In Eq. (\ref{d3}) we have introduced a
strength parameter $d(t)$ with the radial trap. Normally, in the presence
of the radial trap $d(t)=1$. When the radial trap is switched off $d(t)$
will be reduced to 0.

The spherically-symmetric limit of the three-dimensional GP equation
(\ref{d1}) is
obtained by taking $\omega_z= \omega_\rho=\omega_r$. 
In the spherically symmetric configuration, $\Psi({\vec r},
t)=
\varphi(r,t) $. Then the radial part of the GP equation (\ref{d1}) becomes 
\begin{eqnarray}\label{d4}
 \biggr[&-&i\frac{\partial
}{\partial t} -\frac{\partial^2}{\partial
r^2}  -\frac{2}{r}\frac{\partial}{\partial r}           
 +\frac{\omega_r^2}{4\Omega^2} r^2 d(t) 
\nonumber \\ &+& 8\pi n\sqrt {2 } \left|
{\varphi({r};t)}\right|^2 \biggr] \varphi({r};t)=0
 \end{eqnarray}
with  normalization $4\pi \int_0^\infty r^2 dr |\varphi({r};t)|^2=1.$
Here, as in Eq. (\ref{d3}), $d(t)$ is a strength parameter which is to be
reduced to 0 from 1 when the radial trap is switched off. 

Equations (\ref{d3}) and  (\ref{d4}) represent nonrotating ($L=0$) and
vortex solitons ($L\ne 0$) rotating around $z$ axis in the quasi
two-dimensional and spherically-symmetric three-dimensional cases,
respectively. To study the  solitons we  finally set 
$d(t)=0$ in these equations.
It should be noted that $\Omega$ is supposed to be a constant 
reference frequency and not the trap frequencies $\omega_\rho$ 
or $\omega_z$. However, we took 
in Eqs. (3) $-$ (5) the condition $\omega_z=\Omega$. This  does not 
correspond to any specialization but
only simplifies 
the equations algebraically.
Nevertheless, it is of advantage  
to take  $\Omega$  to have the  same
order of
magnitude as  an experimental trap frequency, for example  $\Omega
\equiv 2\pi \times
80$ Hz. With this value of  $\Omega$ the dimensionless time unit
corresponds to  $\Omega^{-1}= 1/(2\pi\times 80)$ s $\approx$ 2 ms.     
In Eq. (\ref{d4}), $\Omega$ is a constant and  $\Omega \ne \omega_r$.

We solve the GP equations (\ref{d3}) and (\ref{d4}) numerically using  the
split-step time-iteration method employing the Crank-Nicholson
discretization
scheme described recently \cite{11}. 
The time iteration is started with the
known oscillator solution of these equations with zero nonlinearity $n$.
Then in the course of time iteration the attractive nonlinearity is 
switched on very slowly  and in the initial stage the harmonic trap is
also switched off 
slowly by changing $d(t)$ from 1 to 0.  If the nonlinearity is
increased rapidly the system collapses. The tendency to
collapse must be avoided to obtain a stabilized soliton. 
After switching off the
harmonic trap 
in Eqs. (\ref{d3}) and
(\ref{d4}) and after slowly introducing a final attractive nonlinearity 
$n_{0}$, 
if  $n$ is replaced by $n_{0}[1-c \sin(\omega t)]$ 
a stabilization of the final solution could be obtained for a suitably
chosen $c$ and a large $\omega$. 
The stabilization could be obtained
for a range of values of $c$ and $\omega$ provided that
$n_{0}$ is negative corresponding to attraction. After some
experimentation with  Eqs. (\ref{d3}) and
(\ref{d4}) we opted for the choice $c=4$ and $\omega=10\pi$ in all our
calculations except in Fig. 1 (b) in 2D and 3D $-$ variational and
numerical. 
In Fig. 1 (b) we report some results for  $\omega=20\pi$ for comparison.

\section{Variational  Results}

To understand how the stabilization can take place 
 we employ a variational method with the following   
Gaussian
wave function for the solution of Eqs. (\ref{d3}) and  (\ref{d4}) 
\cite{ueda,abdul}
\begin{equation}\label{twf}
\varphi(r,t)= N(t)r^L\exp\left[-\frac{r^2}{2R^2(t)}
+\frac{i}{2}{ \beta(t) }r^2+i\alpha(t) 
\right],
\end{equation}
where $N(t)$,  $R(t)$, $\beta(t)$, and $\alpha(t)$ are the
normalization, width, chirp, and
phase of the soliton, respectively.  In 3D
$N(t)=[\pi^{3/4}R^{3/2}(t)]^{-1}$ and $L=0$, and in 2D
$N(t)=[\pi^{1/2}R^{L+1}(t)\sqrt{L!}]^{-1}$.  The Lagrangian density for
generating Eq. (\ref{d4}) with $\omega_r= 0$  is \cite{abdul}
\begin{equation}
{\cal L}(\varphi)=\frac{i}{2}\left(\frac{\partial \varphi}{\partial
t}\varphi^*
- \frac{\partial  \varphi^*}{\partial t} \varphi 
\right)-
\left|\frac{\partial
 \varphi}{\partial r} \right|^2-\frac{L^2| \varphi|^2}{r^2}
-\frac{1}{2} g | \varphi|^4 
, 
\end{equation} 
where $g\equiv 8\pi n \sqrt 2$ in 3D and  $g\equiv 4 n
\sqrt {2\pi}$ in 2D. The trial wave function (\ref{twf}) is
substituted in the Lagrangian density and the 
effective Lagrangian is calculated by
integrating the Lagrangian density: $L_{\mbox{eff}}= \int {\cal
L}(\varphi)
d \vec r.$

The Euler-Lagrange equations for $R(t)$ and $\beta(t)$ are  
then obtained from
the effective Lagrangian in standard fashion in 3D:
\begin{eqnarray}\label{el1}
\frac{dR(t)}{dt}&=& 2 R(t) \beta (t), \\
\frac{d\beta (t)}{dt}&=& \frac{2}{R^4(t)}-2\beta^2 (t)+\frac{g}{2
\sqrt{2\pi^3}
R^5(t)}. \label{el2}
\end{eqnarray}

From Eqs. (\ref{el1}) and  (\ref{el2}) we get the following second-order
differential equation for the evolution of the width
\begin{equation}\label{el}
\frac{d^2 R(t)}{dt^2}=\frac{4}{R^3(t)}+\frac{g_0+g_1\sin (\omega
t)}{\sqrt{2 \pi^3}R^4(t)},
\end{equation}
with
$g=g_0+g_1\sin (\omega   t)$, where $g_0$ corresponds to the constant part
of the scattering length and $g_1$ to the oscillating part. We separate
$R(t)$ into a slowly varying part $A(t)$ and a rapidly varying part $B(t)$
by $R(t)=A(t)+B(t)$. Substituting this into Eq. (\ref{el}) and retaining
terms on the order of $\omega^{-2}$ in $B(t)$ we obtain the following
equations of motion for $B(t)$ and $A(t)$:
\begin{eqnarray}
\frac{d^2 B(t)}{dt^2}&=& \frac{g_1\sin (\omega t)}{\sqrt{2\pi^3}A^4
(t)}\nonumber \\
\frac{d^2 A(t)}{dt^2}&=& \frac{4}{A^3(t)}+\frac{g_0}{\sqrt{2\pi^3}A^4 (t)}
-\frac{2\sqrt 2 g_1 \langle{B(t)\sin(\omega t)}\rangle 
}{\pi^{3/2}A^5(t)},\nonumber
\end{eqnarray}
where $\langle\quad \rangle$ denotes time average over rapid
oscillation.

Using the solution $B(t)=-g_1\sin (\omega
t)/[\sqrt{2\pi^3}\omega^2A^4(t)]$, the equation of motion for $A(t)$
becomes 
\begin{eqnarray}\label{ey}
\frac{d^2 A(t)}{dt^2}&=& \frac{4}{A^3}+\frac{g_0}{\sqrt{2\pi^3}A^4 }
+\frac{g_1^2}{\pi^3\omega^2 A^9}\\
&=& -\frac{\partial }{\partial A}\left[
\frac{2}{A^2}+\frac{g_0}{3\sqrt{2\pi^3 }A^3}+\frac{g_1^2}{8\pi^3\omega^2
A^8} \right].\label{ex}
\end{eqnarray}
The quantity in the square bracket in Eq. (\ref{ex}) is the effective
potential $U(A)$ of the equation of motion:
\begin{eqnarray}\label{eff}
U(A)= \frac{2}{A^2}+\frac{g_0}{3\sqrt{2\pi^3
}A^3}+\frac{g_1^2}{8\pi^3\omega^2
A^8}. 
\end{eqnarray}
Small oscillations around a stable
configuration 
are possible when 
there is a minimum in this effective potential \cite{ueda}. Unfortunately, this
condition
does not lead to a simple analytical solution. 
However, straightforward
numerical study reveals that this effective potential has a minimum 
for a
negative $g_0$  corresponding to attraction
with  
$|g_0|$ above a critical value. For a numerical calculation 
the quantity $g$ is taken to be of the form 
$g=g_0+g_1\sin(\omega t)= g_0[1-4\sin(\omega t)]$ so that
$g_1=-4g_0$ with $g_0$ 
negative (attractive). 

In Figs. 1 (a) and (b) we plot the effective potential 
$U(A)$ vs. $A$
for different $g_0$ for $\omega = 10\pi$ and $\omega = 20\pi$,
respectively. We find that, as the value of $g_0$ is reduced, the
effective
potential develops a minimum which gradually becomes deeper and deeper.  
The depth of the minimum in the effective potential increases as $\omega$
increases. 
For $\omega=10\pi$ and $g_0=-100$ there is no minimum in the effective
potential $U(A)$,
whereas a minimum has  appeared for  $g_0=-200$ which becomes deeper
for $g_0=-300$ and $-500$. For $\omega=20\pi$ and $g_0=-100$, a minimum
has already appeared in Fig. 1 (b). In the rest of this study 
we use  the frequency $\omega = 10\pi$, 
although  its actual value has no consequence on 
the calculation so long as it is  large corresponding to 
rapid oscillation.  
A careful examination reveals that the
threshold for the minimum in the present case is given by
$g_0\approx -116$  for  $\omega = 10\pi$.
Hence, for   $\omega = 10\pi$ stabilization is not
possible for $g_0=-100$, and it is possible for  $g_0<-116.$ 
Also
there is no upper limit for  $|g_0|$ and stabilization is
possible for an arbitrarily large   $|g_0|$. 
As the first
and the third terms on the right hand side (rhs)  of Eq. (\ref{eff}) are
positive, no
stabilization is possible for a positive $g_0$ corresponding to repulsion.
We shall verify these findings by actual numerical calculation in the
following.

\begin{figure}
 
\begin{center}
\includegraphics[width=1.\linewidth]{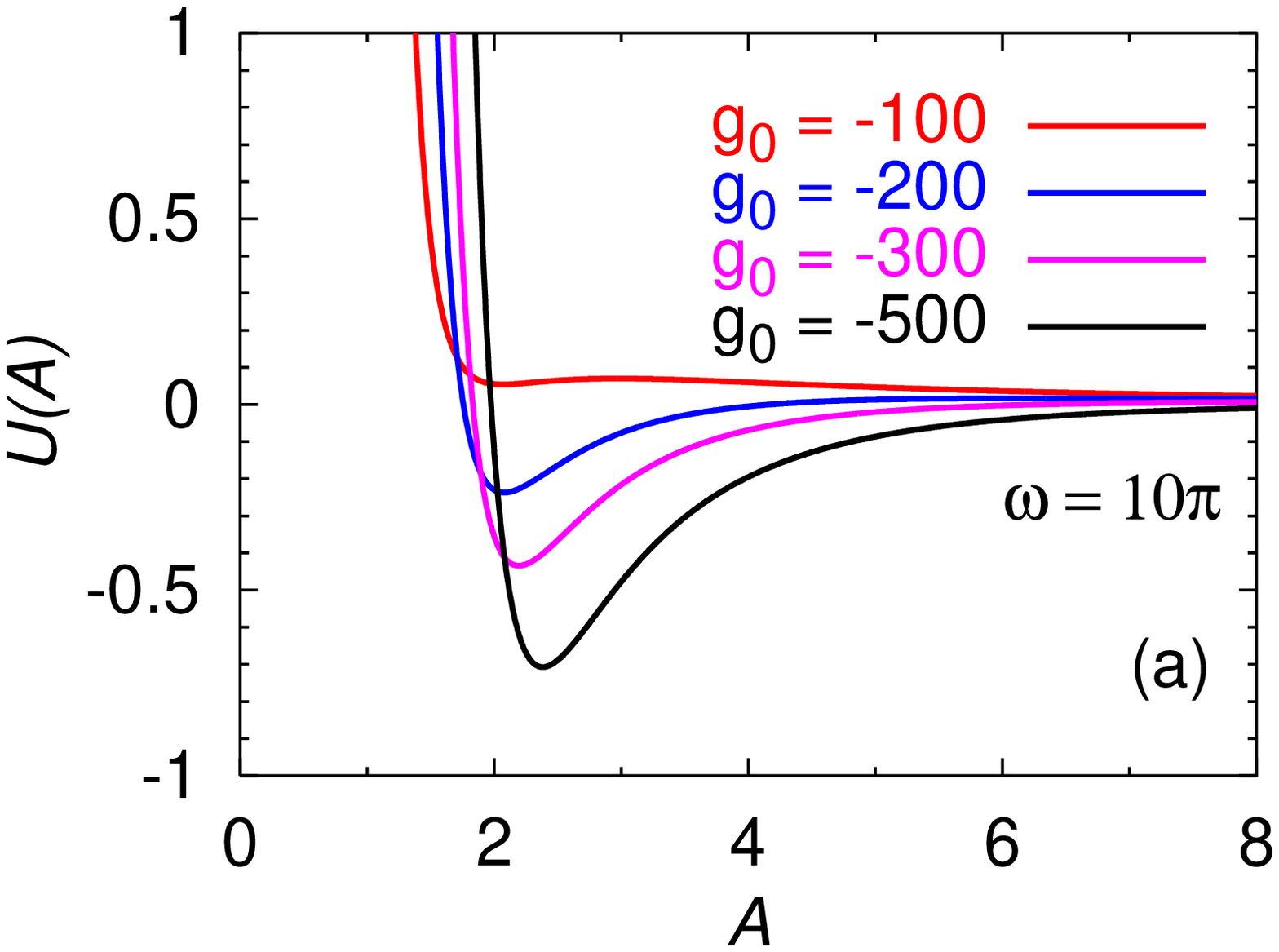}
\includegraphics[width=1.\linewidth]{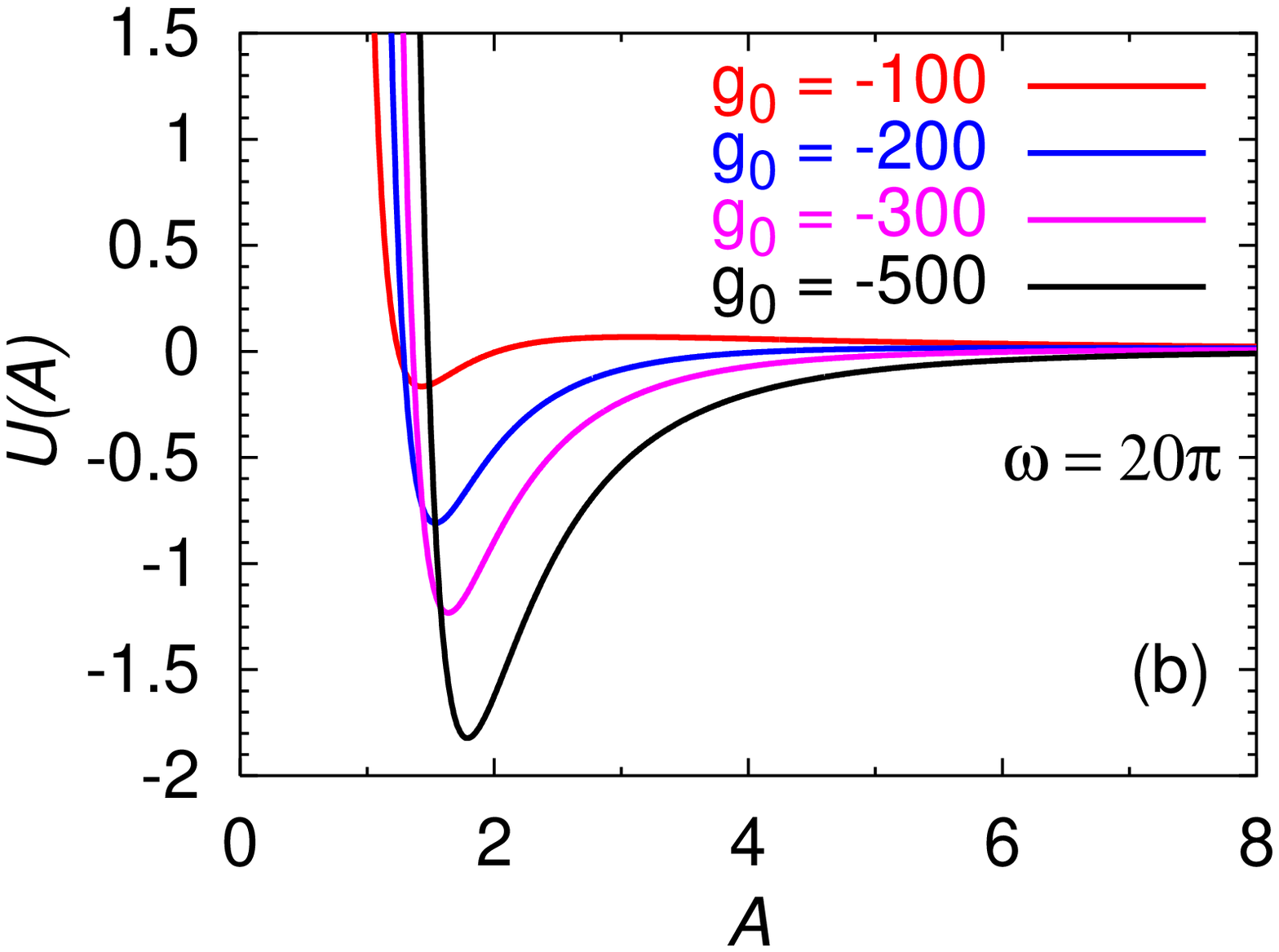}
\end{center}
 
\caption{The effective potential $U(A)$ of Eq. (\ref{eff}) vs. $A$  in
arbitrary units for $g_0=-100,-200, -300,$ and $-500$ (from upper to
lower curve) for (a) $\omega
=10\pi$, and (b) $\omega = 20\pi$.}
\end{figure}

Similarly,  a  two-dimensional soliton of
Eq. (\ref{d3}) leads to the following equation of motion for the small
oscillation of width $ A(t)$ for a general $L$:
\begin{equation}\label{pq}
\frac{d^2 A(t)}{dt^2}=  -\frac{\partial }{\partial A}\left[
\frac{g_0+4\pi}{\pi A^2}+\frac{1}{4}\frac{g_1^2}{\pi^2 \omega^2
A^6}\right].
\end{equation}
A similar equation was obtained before for $L=0$ \cite{ueda}.
 The condition for small oscillations 
is given by $g_0< -4\pi$ or $n<-\sqrt{\pi/2}$, when the first term on the
rhs of   
Eq. (\ref{pq}) becomes negative allowing for the possibility of a minimum
in the effective potential in the square bracket resulting in stable small
oscillations.

\section{Numerical  Results}

With this preliminary variational study we turn to a full numerical
investigation of Eqs. (\ref{d3}) and (\ref{d4}) in 2D and 3D.
As a warm up it is worthwhile to redo the numerical study 
for a $L=0$
soliton in 2D and extend it to a  $L=1$
vortex soliton before considering a $L=0$ soliton in 3D.

\begin{figure}
 
\begin{center}
\includegraphics[width=1.\linewidth]{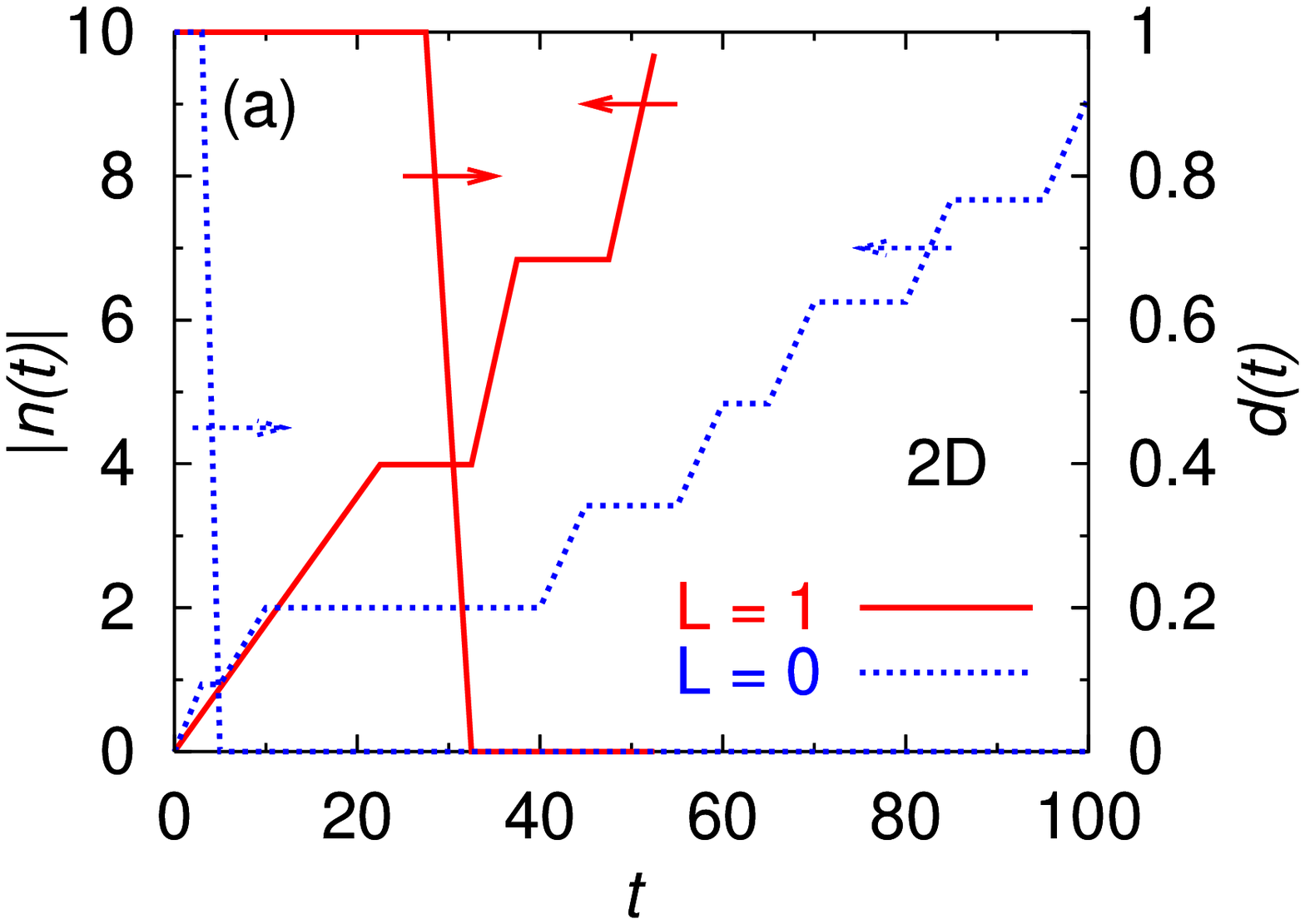}
\includegraphics[width=1.\linewidth]{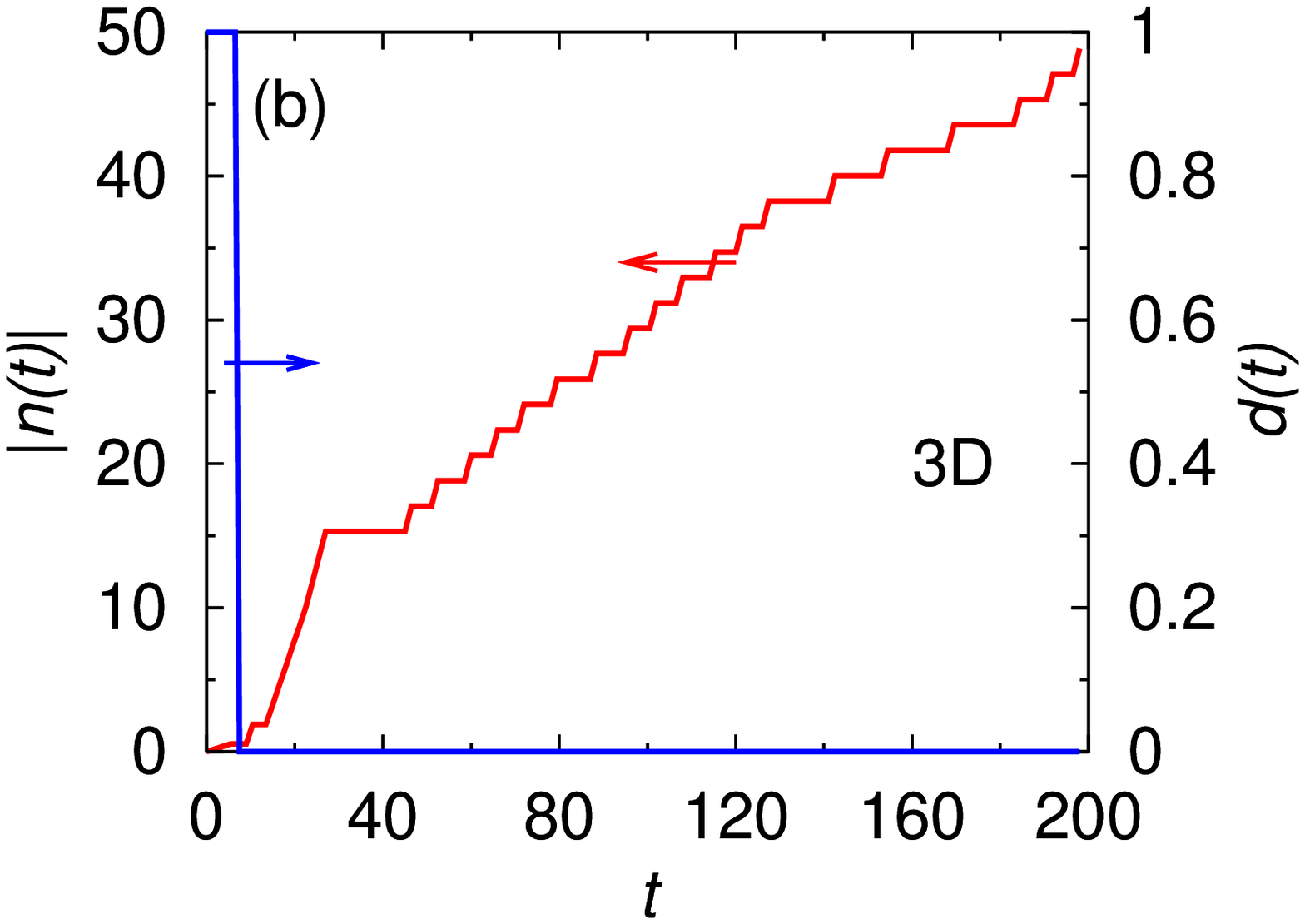}
\end{center}
 
\caption{Variation of nonlinearity parameter $|n(t)|$ and the strength of
radial
trap  $d(t)$ in  (a) Eq. (\ref{d3}) and (b) Eq. (\ref{d4}) in the initial
stage of
stabilization until the desired final nonlinearity $n_0$ is attained at
time $t_0$. For $t>t_0$, the oscillating nonlinearity
$n=n_0[1-4\sin\{10\pi
(t-t_0)\}]$ is applied.  In 2D, $n_0=-9.1$ for $L=0$ and  $n_0=-9.7$ for
$L=1$; in 3D,   $n_0=-48.9$.} \end{figure}

There could be many ways of numerical stabilization of the soliton. In the
course of time evolution of the GP equation certain initial conditions are
necessary for the stabilization of a soliton with a specific nonlinearity
$|n_0|$ above a critical value. As one requires a large (attractive)
nonlinearity $|n_0|$ for stabilization, one needs to reduce the harmonic
trap frequency while increasing the nonlinearity $|n|$. Unless the trap
frequency is reduced the system will collapse \cite{9} due to attraction.
In other words one allows the system to expand and simultaneously
increase the nonlinearity $|n|$. During this process the harmonic trap
is removed, and after the final nonlinearity $n_0$ is attained at time
$t_0$ the
periodically
oscillating nonlinearity $n=n_0[1-4\sin\{10\pi(t-t_0)\}]$ is applied
for $t>t_0$. 
If the size of the condensate is
close to the desired size, a stabilization of the condensate for a large
time is obtained. 
This procedure could also be followed in an experimental
attempt to stabilize a soliton.  Saito {\it et al.} \cite{ueda} used a
qualitatively similar, but quantitatively different procedure for
stabilization. The procedure of Saito {\it et al.} could also be applied
successfully in the present context. 

The correct implementation  of the above
calculational 
scheme
is important  for stabilization.
If the (attractive)
nonlinearity after switching off the harmonic trap is strong for the size
of the condensate
the system becomes highly attractive in the final stage and it eventually
collapses. If the nonlinearity after switching off the harmonic trap 
is weak  for its size
the system becomes weakly  attractive in the final stage  and it expands
to infinity. The final nonlinearity has to have an appropriate
intermediate value, decided by trial,  
for eventual  stabilization. 
In Figs. 2 (a) and (b) we provide the
actual 
time
variation of nonlinearity  $|n(t)|$  as well as the strength parameter for
radial trap $d(t)$ employed 
in Eqs. (\ref{d3}) and (\ref{d4})
for 2D and 3D, respectively. 
A finetuning of the final nonlinearity $|n(t)|$ was needed for
stabilization over a large interval of time as reported in this paper.

\begin{figure}
 
\begin{center}
\includegraphics[width=1.\linewidth]{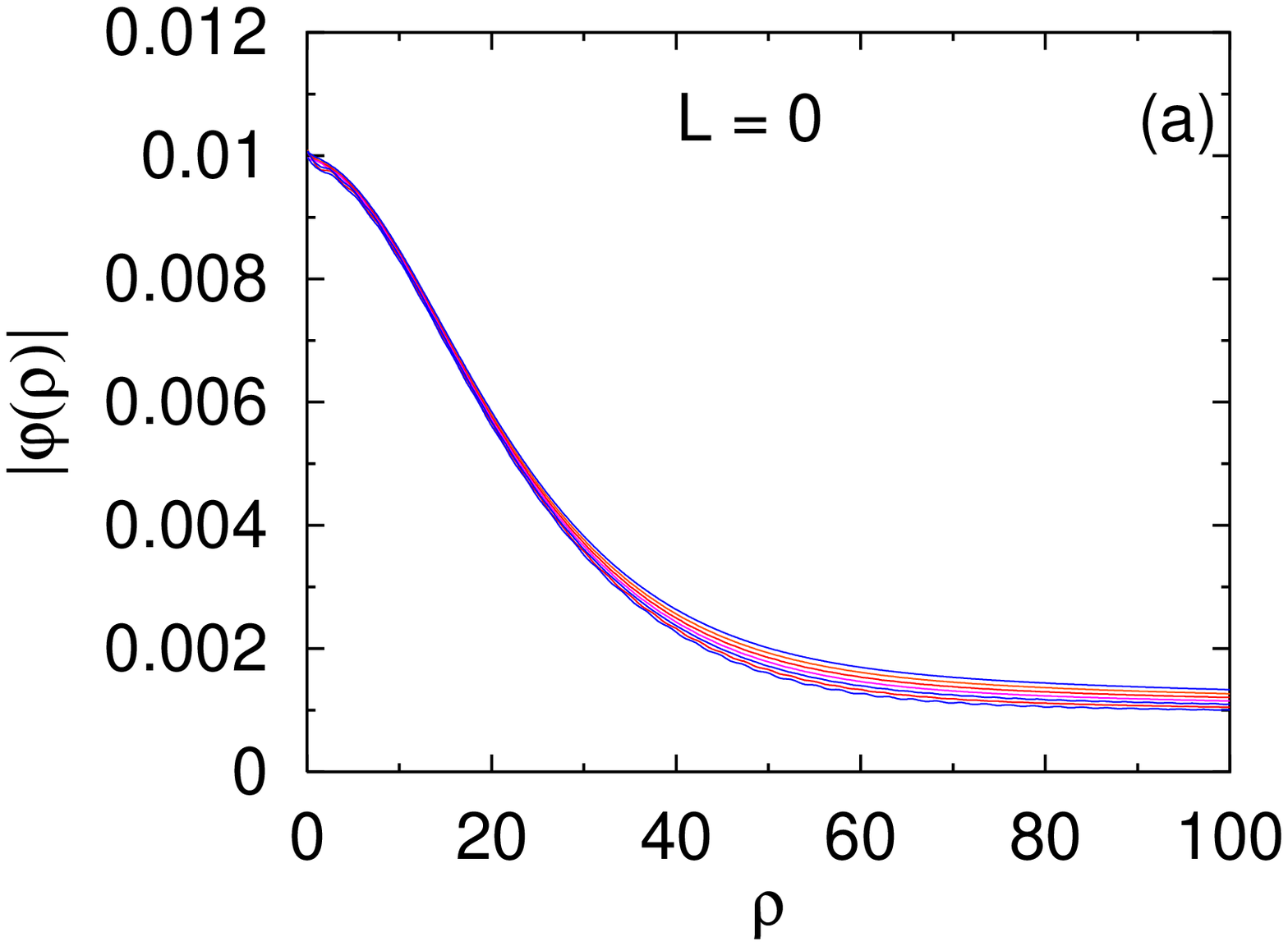}
\includegraphics[width=1.\linewidth]{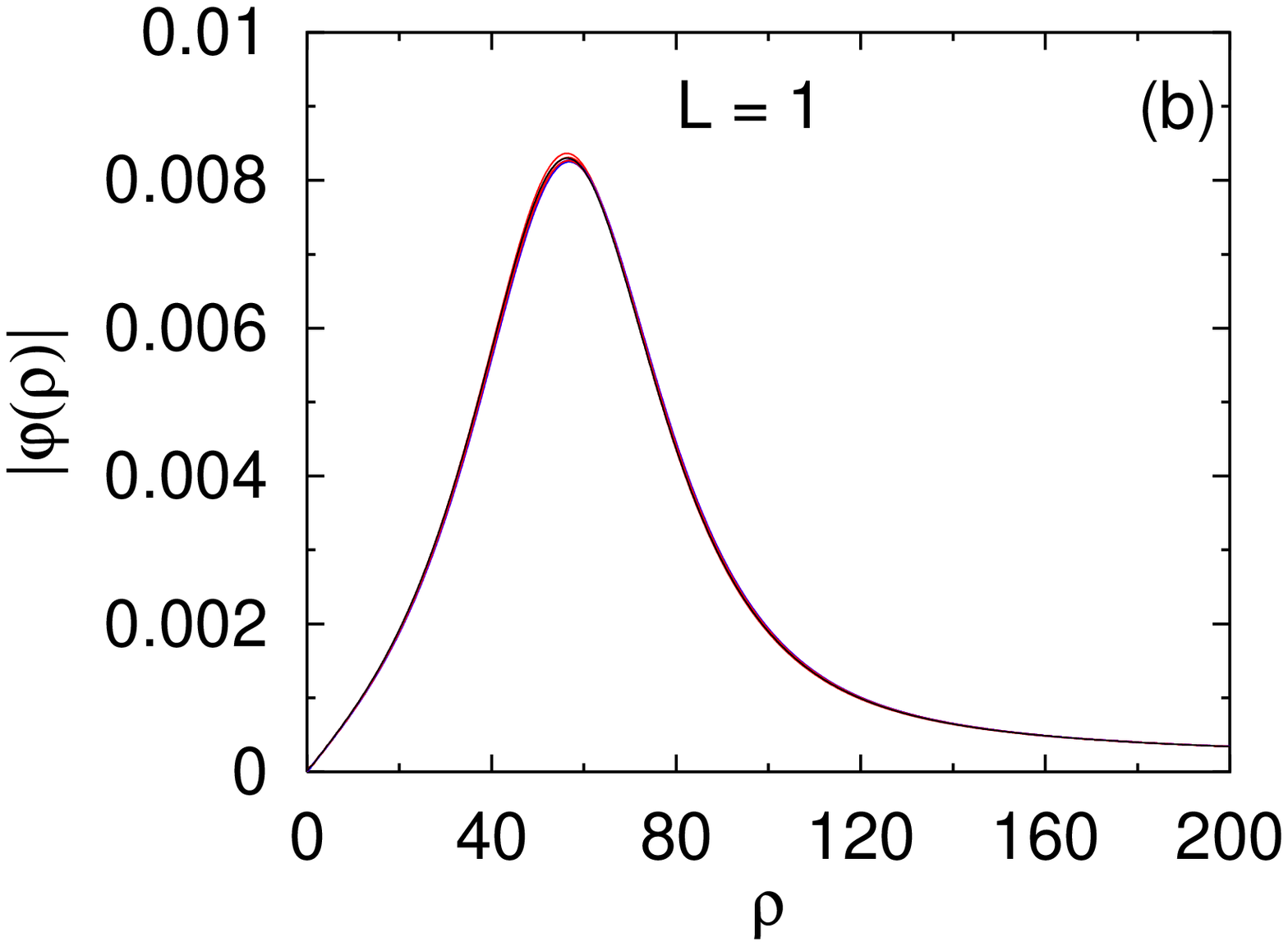}
\end{center}
 
\caption{
(a) Wave function  $|\varphi(\rho)|$ of stabilized soliton in 2D
 with $L=0$ and 
 $n=-9.1[1-4\sin (10\pi t)]$ in Eq. (\ref{d3}) at
times
$t=0,50, 
100,150,200,250,300$.
(b) Same for vortex soliton  with $L=1$ and  
$n=-9.7[1-4\sin (10\pi t)]$ in
Eq. (\ref{d3}) at
times
$t=0,50,100,150,200,250,300$.}
\end{figure}

The results of numerical calculation based on
Eq. (\ref{d3}) are
shown in Figs. 3 (a) and (b) for the two
dimensional soliton ($L=0$) and vortex soliton ($L=1$), respectively,
where we plot the
radial part of the wave function at times $t=0,50,100,150,
200,250,300$ after stabilization is obtained. With the present value of
$\Omega = 2 \pi \times 80$ Hz, 
$t=300$ corresponds to 600 ms.

Next we turn to a numerical calculation in 3D. The relation between the
constant $g$
considered in the variational calculation and nonlinearity $n$ in the GP
equation is $g\equiv 8\pi n \sqrt{2}\approx 35n$. From the
variational calculation presented in Fig. 1 (a) for $\omega=10\pi$,  we
find that
the
condition for stabilization of a soliton  in 3D is  $g_0<-116$ which
corresponds to 
$n_0<-3.3$, approximately. From a complete  numerical solution of the GP
equation (\ref{d4}) we also find that there is a threshold of nonlinearity
for
stabilization consistent
with the variational calculation.
In the numerical calculation it was difficult to obtain accurately the
threshold value of $n_0$ for stabilization. However,  
we could
not obtain stabilization of the soliton for $n_0>-10$. 
The stabilization
is possible for stronger   $n_0$ (larger $|n_0|$).   
In Fig. 4 we plot the radial  wave function of Eq. (\ref{d4})
in 3D at times
$t=0,50,100,150,200,250,300$ for $n_0=-48.9$ after obtaining the
stabilization. 
The narrow spread of the
wave function over
the large
interval of time shows the quality of stabilization.
In both Figs. 3 and 4
the results at intermediate times lie in the region covered by the plots.
The plot of the full wave function, rather than that of 
mean radii or 
the wave function at a particular point vs time, clearly shows the degree
of stabilization achieved.

Of the three cases presented in Figs. 3 and 4
the vortex soliton  of Fig. 3 (b) is  the most stable with minimum
oscillation. For the rotating (vortex) soliton,  the outward
{\it{centrifugal}} force  approximately  balances the attractive
inward
force (as seen in a rotating frame).
The final exact balance is provided by
the oscillating nonlinearity. However, for $L=0$ there is no centrifugal
force  and the stabilization is more difficult.

\begin{figure}
 
\begin{center}
\includegraphics[width=1.\linewidth]{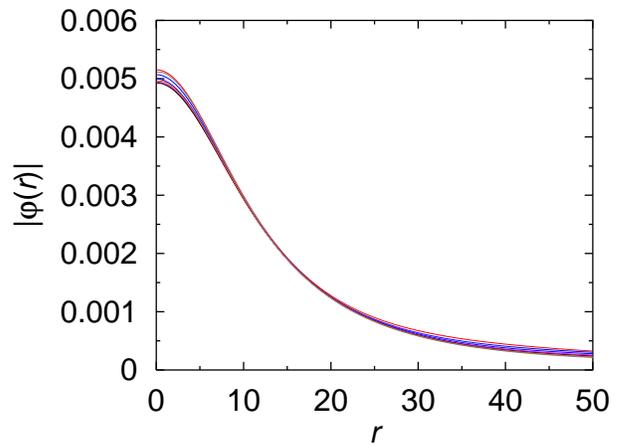}
\end{center}
 
\caption{ Wave function  $|\varphi(r)|$ of stabilized soliton
in 3D
 with 
 $n=-48.9[1-4\sin (10\pi t)]$ in Eq. (\ref{d4}) at
times
$t=0,50,100,150,200,250,300$.}
\end{figure}

The stabilization in both 2D and 3D can be obtained for (attractive) BEC
solitons with nonlinearity $|n_0|$ larger   than a critical value. In 2D
the variational critical nonlinearity is $n_{\mbox{crit}} = -
\sqrt{\pi/2}$,
whereas the final average nonlinearities in Figs. 3 (a) and (b)  are
$n_{0}=-9.1$ and $-9.7$,
respectively. 
In actual numerical calculation we found
that stronger  the nonlinearity $|n_0|$, more sustained 
was the stabilization of 
the soliton. The effective potential develops a deeper minimum for a
larger   nonlinearity $|n_0|$. 
The variational threshold for
stabilization
in 3D is  $|n_{\mbox{crit}}|\approx 3.3$.
In Fig. 4 the actual final average   nonlinearity $|n_0|=48.9$
is much larger than
the variational threshold   $|n_{\mbox{crit}}|\approx 3.3$.

Using a  variational procedure alone, not quite identical with the present
approach,  Abdullaev {\it et al.} \cite{abdul}  also had
found that a stabilization of a soliton could be possible in 3D via a
temporal modulation of the scattering length. However, they confirmed
after further analytical and numerical studies that such a stabilization 
does not take place in 3D. Saito  {\it et al.} \cite{ueda}, on the other
hand, are
silent about the possibility of the stabilization of a soliton in 3D. 
We point out  one possible reason for the negative result obtained 
in 3D \cite{abdul}.
The nonlinearity parameter $\Lambda=-1$  used in 
Ref. \cite{abdul} for stabilizing a soliton in 3D 
corresponds in our notation to $n_{0}=
-\sqrt{2\pi^3}/[8\pi \sqrt 2]= -\sqrt \pi /8\approx -0.22$. [The
relation between $n_0$ and $\Lambda$ of Ref. \cite{abdul} follows
from present Eq. (\ref{el}) and their Eq. (37)]. The nonlinearity
$\Lambda=-1$
is much
too
weak 
for
obtaining a stabilized soliton in 3D. It should be noted that the value of
$n_{0}$ used for the stabilization of a soliton in 3D 
in the present calculation is $-48.9$, whereas the variational threshold
for
stabilization is $n_{\mbox{crit}}\approx -3.3$.  These values of 
nonlinearities  are 
much stronger than the
value
$n_{0}=-0.22$ used in Ref.       \cite{abdul}.

\section{discussion and conclusion}

In this paper we have discussed the stabilization of a bright vortex
soliton in 2D and a bright soliton in 3D by a periodic temporal modulation
of the scattering length.  Now we compare the dimensionless parameters
used in the simulations to typical numbers for an experimental system
$^{85}$Rb. This system has a Feshbach resonance which can be used to vary
the effective interaction between atoms by varying the atomic scattering
length \cite{donley}.  With the reference trap $\Omega = 2\pi \times 80$
Hz,
for $^{85}$Rb the dimensionless length parameter $l=\sqrt{\hbar
/(m\Omega)} \approx 1.2$ $\mu$m.  In a 3D condensate of 10000 atoms, the
variational critical nonlinearity $|n_{\mbox{crit}}|=3.3$ for this system
corresponds to a scattering length $a= n_{\mbox{crit}} l/N\approx -0.4$
nm. The applied nonlinearity $n_0=-48.9$ of the present 3D simulation
corresponds to a scattering length $a\approx -5.9$ nm. The oscillating
nonlinearity corresponds in this case ($a\approx -5.9$ nm) to a variation
of scattering length
between $-30$ nm and 18 nm.  For 100000 atoms the above values for
scattering length will be reduced by a factor of 10. Similar variations of
scattering length of $^{85}$Rb have already been realized in the
laboratory via a Feshbach resonance in actual  experiments
\cite{donley}. Hence it might be possible to stabilize a
$^{85}$Rb condensate using a Feshbach resonance. With 
$\Omega = 2\pi \times 80$ Hz, the interval of stabilization of 300 units
of time in Figs. 3 and 4 corresponds to 600 ms,  which is a reasonably
large interval of time.

The present study has  important consequence in the
generation of a stable 
spatiotemporal soliton  of nonlinear optics  \cite{1,st} which is an
  optical wave packet
confined in all three directions and 
often referred to as a light bullet. The existence and stability of self
trapped beams in a nonlinear medium is a subject of active research since
its suggestion \cite{st}. Such a spatiotemporal optical soliton satisfies
an equation in an anomalously  dispersive medium
quite similar to 
Eq. (\ref{d4}) with $\omega_r = 0$ \cite{1}. Hence a stable solution
for a light bullet in actual 3D can be obtained through a
modulation of the cubic Kerr
nonlinearity. This modulation  can be
achieved in a layered nonlinear  medium with sign-altering 
Kerr nonlinearity 
 \cite{abdul,mal}. 
The possibility of such a stabilization in
two   space dimensions was demonstrated in \cite{mal}, whereas its
impossibility in 3D has been emphasized in
 \cite{abdul}.

In conclusion,  from a numerical solution of the GP equation we find 
that 
it is possible to stabilize a
matter-wave bright soliton in 3D and a vortex soliton in 2D by employing a
rapid
periodic modulation of the scattering length $a$ via a Feshbach resonance
with
an attractive (negative) mean value $a_0$ via: $a \to a_0[1-c \sin(\omega
t)] $ with a large $c$ and $\omega$. From a variational calculation we
show that 
this  oscillation produces a
minimum in the effective potential, thus producing a potential well in
which the soliton can be trapped and execute small oscillations. The
sinusoidal variation of $a$ is actually not needed for stabilization,  any
periodic fluctuation between positive and negative  values stabilizes
the
soliton.  This is
of interest to investigate if such BEC ``bullets" could be created
experimentally in 3D. As the mathematical equation satisfied
by a light bullet \cite{1} in 3D is quite similar to  the
nonlinear three-dimensional equation (\ref{d4})
we suggest the possibility of creating light bullets in a layered Kerr
medium with sign-altering nonlinearity.

\acknowledgments

I thank  Dr. R. A. Kraenkel 
for  informative discussions. 
The work was supported in part by the CNPq 
of Brazil.

\end{document}